\documentclass[prc,aps,twocolumn,superscriptaddress]{revtex4-2}
\usepackage{graphicx,latexsym,amssymb,amsmath,color,multirow,mathrsfs,epsfig,bm}
\usepackage{changes}
\usepackage{hyperref}
\usepackage{longtable}
\usepackage{array}

\hypersetup{colorlinks=true,linkcolor=blue,filecolor=magenta,urlcolor=cyan}

\begin{document}

\title{Long-Lived Opposite-Parity States and the Onset of Octupole Collectivity\\ in Atomic Nuclei} 

    \author{Bui Minh Loc}
	\email{lmbui@sdsu.edu}
    \affiliation{San Diego State University, 5500 Campanile Drive, San Diego, CA 92182}
    \author{Hoang Thai An}
    \affiliation{Department of Physics, Ho Chi Minh City University of Education, 280 An Duong Vuong, Cho Quan Ward, Ho Chi Minh City, Vietnam}
    \author{Nguyen Le Anh}
	\email{anhnl@hcmue.edu.vn}
	\affiliation{Department of Physics, Ho Chi Minh City University of Education, 280 An Duong Vuong, Cho Quan Ward, Ho Chi Minh City, Vietnam}
    \author{Panagiota Papakonstantinou}
	\email{ppapakon@ibs.re.kr}
	\affiliation{Rare Isotope Science Project, Institute for Basic Science, Daejeon 34047, Korea}
 	\author{Naftali Auerbach}
	\email{auerbach@tauex.tau.ac.il}
	\affiliation{School of Physics and Astronomy, Tel Aviv University, Tel Aviv 69978, Israel}

\begin{abstract}
Octupole deformation in atomic nuclei is of interest for both nuclear structure and precision tests of fundamental symmetries, but identifying regions of octupole collectivity remains challenging. We analyze low-energy spectra of odd-mass nuclei and uncover a previously unrecognized empirical regularity that serves as a signature of octupole collectivity in neighboring even-even systems. The observed patterns, which can be understood within a core-coupling picture, are consistent with previous theoretical studies and lead to predictions for neutron-rich and proton-deficient nuclei. These findings provide a simple empirical guide for identifying promising candidates for future experiments and microscopic calculations.

\begin{center}
    This work was inspired by the Nobel Lecture of Maria Göppert Mayer \\ and is dedicated to the 120th anniversary of her birth (June 28, 1906).
\end{center}

\end{abstract}

\maketitle

{\em Introduction---}
Since the mid-1990s, the study of octupole deformation in nuclear physics has gained more attention after proposals to use octupole-deformed nuclei in electric dipole moment experiments \cite{AuerbachPRL1996, SpevakPRC1997}. In contrast to quadrupole deformation, octupole deformation is weak and confined to limited regions of the nuclear chart. To this day, knowing whether a nucleus exhibits octupole deformation in its ground state remains a challenging problem. Compelling evidence for octupole deformation requires a coherent interpretation of multiple experimental observables supported by theoretical modeling (see Ref.~\cite{Butler2020Review} for the most recent review).

The study of octupole deformation has a long history dating back to the early days of nuclear physics \cite{LaneNP1960}. 
Nowadays, octupole correlations are investigated using a variety of theoretical approaches, including shell-model calculations \cite{BouhelalPRC962017, YoshinagaPRC1092024}, non-relativistic (see Ref.~\cite{ButlerJPG432016} for a review) and relativistic self-consistent mean-field theory, and beyond \cite{AgbemavaPRC932016, XiaPRC962017, RongPLB8402023}, interacting-boson models~\cite{NomuraPRC892014}, generator coordinate method \cite{ZhouIJMPE2023}, and many others. Each theoretical model, however, encounters its own challenges. 

Self-consistent mean-field theory offers a consistent approach to octupole excitations throughout the nuclear chart.
In contrast to the shell model, cross-shell excitations are incorporated naturally, making these approaches particularly successful in identifying regions of octupole softness and octupole deformation across the nuclear chart \cite{RobledoPRC842011, LocPRC2023}. Nevertheless, the interpretation of octupole correlations remains challenging. 
In particular, nuclei near the transition between octupole softness and static octupole deformation exhibit shallow energy surfaces, making theoretical predictions sensitive to the choice of energy density functional. The long-standing difficulty in reproducing the unusually large low-lying octupole excitation in $^{96}$Zr provides a representative example of this challenge \cite{LocPRC2023}.
In general, it remains difficult to determine unambiguously whether a nucleus should be classified as octupole-soft or statically octupole-deformed.

This limitation motivates us to search for a complementary observables-based method. Historically, some of the most important advances in nuclear physics have emerged from systematic patterns in experimental data. A notable example is the work of Maria G\"oppert Mayer, who recognized shell closures through systematic trends in nuclear masses, abundances, and excitation spectra before their microscopic origin was fully understood. The subsequent development of the nuclear shell model by Mayer and independently by Haxel, Jensen, and Suess provided a theoretical explanation for these observations \cite{MayerPR751949, HaxelPR751949}.

A similar strategy has proven fruitful in the study of octupole correlations. Through a systematic analysis of octupole transition strengths across the nuclear chart, Spear and Catford \cite{SpearPRC411990} and Cottle \cite{CottlePRC421990} identified the octupole magic numbers, which characterize octupole-soft nuclei. 
Inspired by these examples, we seek a complementary observables-based signature that can distinguish octupole-soft nuclei from those with static octupole deformation. 
We identify the onset of shape isomerism in odd-parity odd-mass nuclei as a signature and offer a physical interpretation of its relevance.

Notably, isomerism associated with octupole transitions already appeared in the work of Maria Göppert Mayer. In her 1963 Nobel Lecture \cite{Mayer1963}, her discussion of shell-orbital effects in isomeric states foreshadowed what later became the concept of octupole magic numbers \cite{NazarewiczNPA429NPA, SpearPRC411990, CottlePRC421990}. 
Her observation of isomeric states in odd-mass nuclei provided an early indication of a possible method to identify static octupole-deformed nuclei.

Nuclear excited states typically decay in picoseconds ($10^{-12}$ s). Excited states with a lifetime of nanoseconds ($10^{-9}$ s) can be considered isomers. Nowadays, isomers are classified into various types \cite{GargADNDT2023} such as shape-, spin-, or $K$-trap isomers, depending on whether decay is hindered by changes in deformation \cite{PolikanovSPU1968, MollerADNDT2012}, total angular momentum \cite{AuerbachPL1964, AuerbachPL1964_2}, or the angular-momentum projection onto the symmetry axis \cite{AuerbachPRC2014}, respectively.
The present work concerns long-lived states and isomeric behavior associated with octupole-induced shape trapping.

In a recent work~\cite{LocPRC2023}, we mapped the landscape of octupole deformation softness across the nuclide chart 
by means of QRPA calculations. For that purpose, we took advantage of the method's formal properties, namely that extremely low-energy or imaginary-energy states in a spherical-basis calculation of excitations signify softness or instability against deformation in the respective natural-parity channel~\cite{Tho1961}. 
In the present work, we adopt what Maria Goeppert Mayer termed the ``experimentalist's approach'', taking advantage of the wealth of spectroscopic data accumulated since the early days of the shell model. 
Guided by modern nuclear data compilations, particularly the National Nuclear Data Center’s NuDat3~\cite{nndc}, and simple structural arguments,  we aim to map the transitional boundary between non-octupole and statically octupole-deformed nuclei.  

While octupole correlations are commonly studied through low-lying collective excitations in even-even nuclei, in the present work we gain additional insight by examining the neighboring odd-mass systems.
In the soft-octupole region, odd-mass nuclei may exhibit long-lived opposite-parity states arising from a mismatch between the intrinsic structures of the ground and octupole-excited configurations, leading to octupole-induced shape trapping and hindered transitions.
Along isotopic or isotonic chains, the gradual appearance and then disappearance of these long-lived states signal the onset of stable octupole deformation in the ground state, as the intrinsic structures of the two configurations become increasingly aligned.

\begin{figure*}[]
    \centering
    \includegraphics[width=0.75\textwidth]{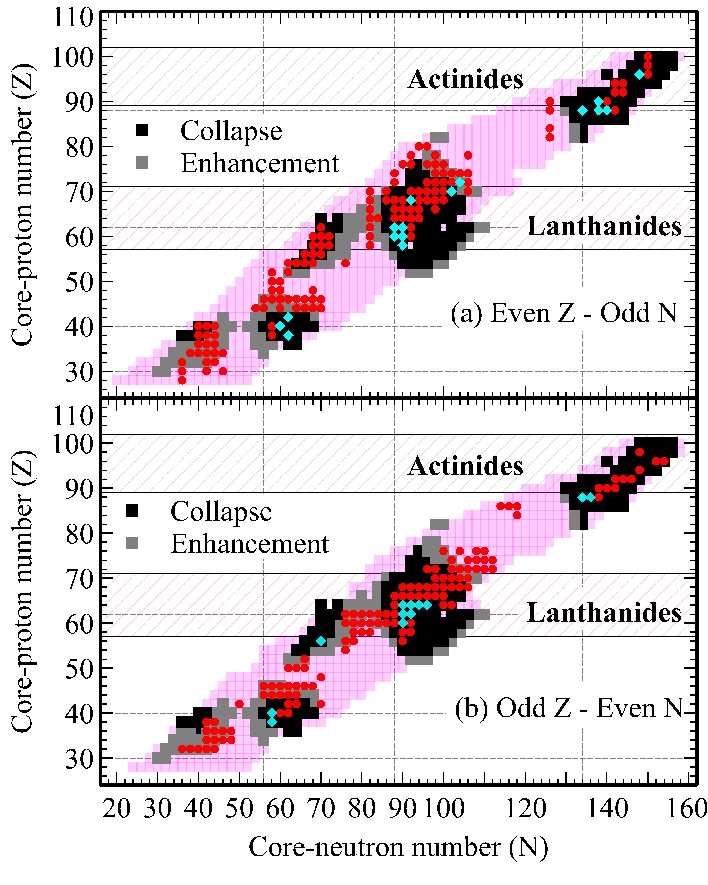}
    \caption{ The dotted lines mark the octupole driving (or magic) numbers. Following Ref.~\cite{SpearPRC411990}, they are 34, 56, 88, and 134 for neutrons and 30, 40, 62, 88 for protons. 
    Octupole soft nuclei are black and gray squares on top of the pink squares, which present nuclei with data available. 
    The proposed octupole-induced shape trappings are red circles. Candidates for statically octupole-deformed nuclei (cyan diamonds) appear inside the black or gray regions and are spatially surrounded by red circles.
    }
    \label{fig:OctTrapSurvey}
\end{figure*}

{\em Spectroscopic patterns---} 
For each even-even nucleus, we inspect the low-lying spectra of adjacent odd-nuclei with one extra neutron or proton provided that the data exist. 
The physical motivation and interpretation will be provided later.  In total, we observe the low-lying spectroscopy of 470 odd-mass nuclei with data available from the National Nuclear Data Center’s NuDat3 \cite{nndc}. 
They are shown as pink squares in Figure~\ref{fig:OctTrapSurvey}. We display the data in the $(Z, N)$ plane
: Figure~\ref{fig:OctTrapSurvey}~(a) displays odd-$N$ nuclei and Figure~\ref{fig:OctTrapSurvey}~(b) displays odd-$Z$ nuclei.

Next, we mark the octupole magic numbers, 34, 56, 88, and 134 for neutrons and 30, 40, 62, and 88 for protons, following Ref.~\cite{SpearPRC411990}, with dashed lines. 
In addition, we show the spherical QRPA results with the SkM$^*$ force \cite{BartelNPA1982} for the even-even nuclei. 
Specifically, we indicate 
the regions where spherical QRPA results are collapsed (the energy of the $3^-_1$ is imaginary) or the $B(E3)$ is strongly enhanced, as was discussed in detail in Ref.~\cite{LocPRC2023}. 
We call such regions, which correspond to the black and gray points in Figure~\ref{fig:OctTrapSurvey}, the octupole-soft regions. 
Red points mark isotopes which display long-lived opposite-parity states ($t_{1/2} > 1$ ns). Cyan points highlight nuclei within the octupole-soft regions which do not display long-lived opposite-parity states. Their significance will be clarified below.

We observe that the presence of long-lived opposite-parity excited states in the experimental data is aligned with the octupole-soft regions predicted by theory. 
The cyan diamonds indicate the absence of long-lived opposite-parity excited states in nuclei that are embedded well within the regions of strong octupole correlations. 
Next, we interpret the long-lived opposite-parity states (red points) as the result of shape trapping within regions of strong octupole correlations and the embedded nuclei with no long-lived opposite-parity states (cyan points) as candidates for static octupole deformation. 
We demonstrate the underlying microscopic mechanism for this pattern by examining the Zr and Ra isotopes as examples. 

{\em Mechanism of Shape Trapping---}  
In even-even nuclei, low-lying octupole states commonly exhibit lifetimes on the nanosecond scale because electric-octupole transitions are intrinsically slow and their lifetimes scale as $E_\gamma^{-7}$. Consequently, a long lifetime alone does not necessarily indicate an unusual nuclear structure.

The situation is different in odd-mass nuclei. Opposite-parity states might have access to electric-dipole ($E1$) decay. One example is the $E1$ transition from $5/2^-_1$ to $3/2^+_{\text{g.s.}}$ in $^{101}$Zr. Both $E1$ and $E3$ transitions are allowed, but the $E1$ transition is much faster than an $E3$ transition as the lifetime $\tau (E1) \propto E_\gamma^{-3}$, whereas $\tau(E3) \propto E_\gamma^{-7}$. 
One therefore expects such low-lying opposite-parity states to decay promptly. When such a state instead exhibits a long lifetime, the $E1$ decay must be strongly hindered by an unusual nuclear structure.

Long-lived nuclear states can arise from a mismatch between the structural wave functions of the initial and final states, which suppresses the transition probability despite the absence of strict selection-rule prohibitions~\cite{PalitEPJST2024}. In the well-known shape isomers, this hindrance is commonly associated with a reduced overlap between wave functions localized in different regions of the potential-energy surface~\cite{MollerADNDT2012}. In the present work, we propose that a similar mechanism operates in octupole-soft nuclei, where the transition connects non-octupole and octupole-correlated configurations in neighboring odd-mass nuclei. 
The mismatch can arise from the different polarization of the soft even-even core by the odd particle. In such a case,  the excited state is associated with an octupole-shaped configuration, whereas the ground state is associated with a non-octupole configuration. Even if the $E1$ transition is allowed by the angular-momentum and parity selection rules, the electric dipole operator cannot efficiently transform an octupole-shaped many-body configuration into a non-octupole one, and the $E1$ matrix element is suppressed. 
The $E1$ transition is allowed but hindered, and the excitation becomes trapped by its shape. 

As octupole correlations develop along an isotopic or isotonic chain, the shape of the even-even core can evolve from a non-octupole shape, resulting in no shape mismatch, through an easily polarized one, potentially leading to a shape mismatch, and toward an octupole one, where there is no shape mismatch between the ground and excited states. Further along the isotopic or isotonic chain, as octupole correlations weaken again, the shape of the even-even core becomes non-octupole again. 
Indeed, we observe in Figure~\ref{fig:OctTrapSurvey} that candidates for octupole deformed nuclei (cyan diamonds) are surrounded, at least partially, by nuclei identified by the appearance of long-lived states in their spectroscopy (red circles). This happens inside the octupole soft region (black and gray).

Next, we demonstrate this proposed mechanism by inspecting more closely the representative examples of the Zr and Ra isotopic chains.

{\em Zr and Ra isotopes---} 
We begin with the Zr isotopic chain 
where the microscopic origin of octupole correlations is particularly transparent \cite{AbbasNPA1981, LocPRC2023}. 
In heavy nuclei, strong octupole correlations typically emerge from the coherent action of several opposite-parity pairs~\cite{LocPRC2023}.
In Zr, on the other hand, the major source of octupole correlations is the opposite-parity neutron pair $d_{5/2}$-$h_{11/2}$, with the additional contribution from the pair $g_{7/2}$-$h_{11/2}$ ~\cite{AbbasNPA1981, LocPRC2023}. 

$^{96}\mathrm{Zr}$ is spherical in its ground state, yet it can be easily polarized toward octupole deformation \cite{LocPRC2023, AbbasNPA1981}. 
It is therefore an octupole-soft nucleus. With the addition of two neutrons, $^{98}\mathrm{Zr}$ is expected to exhibit a stronger tendency toward octupole deformation than $^{96}\mathrm{Zr}$.
Examining the excitation spectrum of the odd-mass nucleus $^{99}\mathrm{Zr}$ with $J^\pi_{\rm{g.s.}} = 1/2^+$, we find excited states $5/2^-$ at 667.48 keV and $7/2^-$ at 678.55 keV. 
Following the particle-core picture \cite{AuerbachPLB1968}, those can be interpreted as $|^{98}\mathrm{Zr}(3^-) \otimes 1/2^+; 5/2^- \rangle$ and $|^{98}\mathrm{Zr}(3^-) \otimes 1/2^+; 7/2^- \rangle$ configurations. 
Both have long lifetimes of $2.6$~ns and $8.9$~ns, respectively, indicating that their shapes are different from those below, including the ground state. 
In addition, the decay branches to the ground state are absent. 
These observations are compatible with the scenario in which the decay of octupole-shape excited states to a non-octupole ground state is strongly hindered.

With the addition of two neutrons, $^{100}\mathrm{Zr}$ should exhibit a stronger tendency toward octupole deformation than $^{98}\mathrm{Zr}$. 
In the odd-mass nucleus $^{101}\mathrm{Zr}$ with $J^\pi_{\rm{g.s.}} = 3/2^+$, the excited $5/2^-_1$ and $7/2^-_1$ states at $216.67$~keV and $321.11$~keV have lifetimes of $0.33$~ns and $0.27$~ns, respectively. 
The presence of a strong $5/2^-_1 \rightarrow 3/2^+_{\mathrm{g.s.}}$ decay indicates similar shapes for the $^{100}\mathrm{Zr}$ core in both the excited and the ground states.

In the next odd-mass nucleus $^{103}\mathrm{Zr}$, the ground state is $5/2^-$. The octupole-excited states $3/2^+$ and $5/2^+$ at $258.9$ keV and $357.1$ keV exhibit strong decay to the ground state, respectively. 
This behavior indicates a further development of octupole correlations along the Zr isotopic chain, consistent with a transition to static octupole deformation in $^{102}\mathrm{Zr}$.

By adding more and more nucleons, octupole correlations should weaken again. 
There are no data for heavier isotopes at present, but a prediction of the present approach is that, in heavier Zr isotopes, we would reach a transitional region of soft-octupole nuclei again and eventually non-octupole Zr nuclei. 
This would be in line with the predictions of the QRPA approach~\cite{LocPRC2023}. 
The experimental signature would be a return to fast $E1$ transitions in heavier isotopes, passing through a region of hindered, long-lived $E1$ transitions.  

In summary, the systematics along the Zr isotopic chain suggest a characteristic pattern marked by the appearance and disappearance of long-lived opposite-parity states. 
This behavior is consistent with a transition from non-octupole ground states toward the onset of static octupole deformation with increasing neutron number.

A useful benchmark is the Ra isotopic chain \cite{ButlerPRL2020}. 
Existing data, which point to strong octupole correlations,  are consistent with our interpretation.
First, for $^{220}\mathrm{Ra}$, we consider the neighboring odd-mass nucleus $^{221}\mathrm{Ra}$. The $5/2^-$ and $7/2^-$ octupole-excited states are not long-lived and show strong decay to the ground state. Within the present framework, this behavior is consistent with a reduced structural mismatch between the corresponding intrinsic states and therefore with a static octupole deformation in $^{220}$Ra, in agreement with \cite{GaffneyNature2013}. 

For $^{222}\mathrm{Ra}$, we examine the level structure of the neighboring odd-mass nucleus $^{223}\mathrm{Ra}$, whose ground state has $J^\pi_{\mathrm{g.s.}}=3/2^+$. 
The opposite-parity states with $J^\pi = 3/2^-, 5/2^-$ have lifetimes $0.63$ and $0.24$~ns, respectively, with a dipole transition to the ground state. The $7/2^-$, and $9/2^-$ have measured lifetimes of $0.45$ and $0.20$~ns, respectively, but these two states cannot have a dipole transition to the ground state because of the total angular momentum selection rule. 
The strong decays of the $^{223}$Ra $J^\pi=3/2^-, 5/2^-, 7/2^-$, and $9/2^-$ states to the $J^\pi_{\mathrm{g.s.}}=3/2^+$ ground state are consistent with a small structural mismatch and strong octupole correlations in $^{223}\mathrm{Ra}$. 
 
In the case of $^{225}$Ra, although no direct transition from the opposite-parity excited states to the ground state is observed, this absence does not necessarily imply a shape mismatch between the excited states and the ground state. 
The reason is that $E1$ transitions are possible to intermediate states.
The decay proceeds predominantly through an $E1$ transition to the low-lying $J^\pi = 5/2^+$ state at $25.41$~keV. 
An alternative interpretation is that both excited states display shape mismatch with the ground state. 
Even though the present approach is inconclusive as to what the underlying shape is in the case of $^{224,225}\mathrm{Ra}$, it does not contradict the findings of \cite{GaffneyNature2013} regarding octupole deformation in $^{225}\mathrm{Ra}$.

Similarly, $E1$ transitions are favored in the cases of $^{227,229}$Ra. 
However,  the $5/2^- \rightarrow 5/2^+_{\mathrm{g.s.}}$ transition in $^{229}\mathrm{Ra}$ has a lifetime of $0.66$~ns, compared to $0.236$~ns for the $5/2^- \rightarrow 3/2^+_{\mathrm{g.s.}}$ transition in $^{227}\mathrm{Ra}$. The longer lifetime indicates increased hindrance, suggesting that octupole collectivity in $^{228}\mathrm{Ra}$ is less developed than in $^{226}\mathrm{Ra}$. This interpretation is consistent with Ref.~\cite{ButlerPRL2020}, which concludes that $^{228}\mathrm{Ra}$ exhibits vibrational octupole character.

In summary, the Ra isotopic chain exhibits a systematic evolution consistent with the present framework. The absence of isomeric states in $^{221,223,225}\mathrm{Ra}$ reflects closely aligned intrinsic states and well-developed octupole correlations in the corresponding even-even cores. In contrast, the increased hindrance observed toward $^{229}\mathrm{Ra}$ indicates a reduction of octupole collectivity and possibly a return to non-octupole shapes in heavier isotopes. 
Similarly, a return to weakened octupole correlations in neutron-deficient Ra isotopes would be consistent with the present framework and could be inferred from future spectroscopic data.

{\em Hexadecapole Shape Trapping---} 
The octupole case discussed above involves hindered electric-dipole transitions between states whose structure is influenced by octupole correlations. By analogy, one may ask whether enhanced hexadecapole correlations could manifest through hindered electric-quadrupole transitions. As an illustrative example, enhanced hexadecapole collectivity has been predicted in the vicinity of the $^{122}\mathrm{Te}$ isotopes \cite{AnhPRC2024}. Interestingly, odd-mass Xe isotopes exhibit a pattern reminiscent of the octupole-induced transition hindrance discussed above: In $^{125}\mathrm{Xe}$, a long-lived $J^\pi = 7/2^+$ state is observed with a lifetime of 140 ns, while in $^{127}\mathrm{Xe}$ the corresponding state has a shorter lifetime of 36.7 ns. This isomeric behavior is no longer observed in the heavier isotopes $^{131,133,135}\mathrm{Xe}$.

Although the available evidence is considerably more limited than in the octupole case, the observed lifetime evolution suggests a structural transition associated with enhanced hexadecapole correlations and raises the possibility that the approach proposed here can be applicable beyond octupole collectivity.

{\em Conclusion and Perspectives---} 
We identify octupole-induced shape trapping as the mechanism responsible for the systematic appearance of long-lived opposite-parity states in neighboring odd-mass nuclei. The resulting systematics provide a new experimental signature of the nature of octupole correlations, distinguishing collective octupole vibrations, octupole-soft nuclei, and static octupole deformation.

We interpret this behavior as arising microscopically from octupole-core polarization: long-lived states emerge when the connected states have distinct intrinsic structures and disappear as those structures become increasingly similar. This mechanism makes clear predictions for the occurrence or disappearance of long-lived opposite-parity states in more exotic nuclei, providing a target for future spectroscopic and lifetime measurements as well as quantitative microscopic calculations.

\section*{Acknowledgments}

Supported by the Institute for Basic Science (IBS) through
the National Research Foundation (NRF) of Korea (2013M7A1A1075764) and by the NRF (TOPTIER, RS-2024-00436392).

\bibliographystyle{apsrev4-2}
\bibliography{ref}

\end{document}